# Structure-property relations characterizing the devitrification of Ni-Zr glassy alloy thin films

Debarati Bhattacharya, S. Rayaprol, Kawsar Ali, T. V. Chandrasekhar Rao, P. S. R. Krishna, R. B. Tokas, S. Singh, C. L. Prajapat, and A. Arya

View Online

Export Citation

CrossMark

# Structure-property relations characterizing the devitrification of Ni-Zr glassy alloy thin films

View Online

Export Citation

CrossMark

Debarati Bhattacharya,[1,2,a)] S. Rayaprol,[3] Kawsar Ali,[2,4] T. V. Chandrasekhar Rao,[2,5] P. S. R. Krishna,[1,2] R. B. Tokas,[6] S. Singh,[1,2] C. L. Prajapat,[5] and A. Arya[2,4]

**AFFILIATIONS**

[1]Solid State Physics Division, Bhabha Atomic Research Centre, Mumbai 400085, India
[2]Homi Bhabha National Institute, Anushakti Nagar, Mumbai 400094, India
[3]UGC-DAE Consortium for Scientific Research—Mumbai Centre, Trombay, Mumbai 400085, India
[4]Materials Science Division, Bhabha Atomic Research Centre, Mumbai 400085, India
[5]Technical Physics Division, Bhabha Atomic Research Centre, Mumbai 400085, India
[6]Atomic and Molecular Physics Division, Bhabha Atomic Research Centre, Mumbai 400085, India

a)**Author to whom correspondence should be addressed:** debarati@barc.gov.in

**ABSTRACT**

The investigation of devitrification in thermally annealed nanodimensional glassy alloy thin films provides a comprehensive understanding of their thermal stability, which can be used to explore potential applications. The amorphous to crystalline polymorphous transformation of cosputtered Ni-Zr alloy (Ni78Zr22 at. %) films, with a thickness lower than the reported critical limit of devitrification, was studied through detailed structural characterization and molecular dynamics (MD) simulations. Devitrification to a nanocrystalline state (Ni$_5$Zr$_2$ structure) was observed at 800 °C, with an increase in density (∼3.6%) much higher than that achieved in bulk alloys. Variation in the magnetic property of the films and the overall physical structure including morphology and composition were examined before and after annealing. MD simulations were employed to effectively elucidate not only the high densification but also the increased magnetic moment after annealing, which was correlated with the simulated change in the coordination number around Ni atoms. The structural relaxation process accompanying devitrification was described as a disorder-to-order transformation while highlighting the crucial role played by chemical short range order prevalent in glassy materials.

## I. INTRODUCTION

In an earlier study,[1] it has been established that various alloy compositions can be formed by controlling sputter rates of Ni and Zr targets during the DC magnetron cosputtering process. In the case of as-deposited Ni$_{10}$Zr$_7$ alloy thin films, it was concluded that crystallinity could be achieved as well. We now demonstrate through the present study the cosputtered deposition of very thin amorphous alloy films with an atomic composition Ni78Zr22 (at. %) and report the effect of thermal annealing to produce devitrification through a polymorphous transformation.

Amorphous (glassy) metal alloys[2] are functionally important materials possessing potential applications in various fields.[3] In particular, transition metal glassy amorphous alloys have several properties which are very useful in various commercial applications.

For example, these alloys are utilized in semiconductor technology due to their interesting electronic properties: for chemisorption since the alloy constituents exhibit very different adsorption characteristics, electrocatalytic activity,[4] and in hydrogen absorption. A combination of a small late transition metal (groups VIIIb, Ib: TL) with a large early transition metal (groups IIIb to VIb: TE) results in alloys formed in a wide composition range. The Ni-Zr system is a specific example of this TL-TE group. The constituents react strongly with each other as a consequence of associated highly negative enthalpies of mixing. Properties of these alloy materials are further enhanced when investigated in the thin film form. Thin film metallic glasses (TFMGs) are a well studied class of materials[5] typified by properties superior to their bulk counterparts, i.e., glassy bulk metal alloys. TFMG preparation methods foster thermodynamic conditions favoring the formation of the amorphous state, similar to that of bulk glassy alloys formed through rapid quenching from the liquid state.[6] However, in the case of TFMG formation, the liquid state is completely bypassed. Sputter deposition is the most popular preparation technique for TFMGs. The very high effective quenching rates can promote kinetic conditions with a low value of driving thermodynamic potential for crystallization that can inhibit nucleation and crystallization. The rapidly quenched amorphous films would thus possess a high degree of structural disorder.

When exposed to different thermal treatments, glassy alloy films can undergo devitrification or crystallization through structural relaxation. The resultant films are invariably homogeneous, uniform, and exhibit a very fine (nano) structure.[7] The advantage of forming such crystallized alloy films from the original vitrified state is well elucidated in the literature,[8] particularly with regard to improved mechanical[9] and magnetic[10] properties. The applicability of TFMGs in various fields would thus prompt an interest in gaining better insight into the methodology of the amorphous to crystalline phase transformation, in order to be able to tune desired properties of these materials. Understanding the structure-property relationship in metallic glasses undergoing devitrification hinges on appropriate modeling of local structure. Molecular dynamics (MD) simulations can be effectively employed to relate structural models to the properties of such systems in order to establish their validity. A mechanism of crystallization of metallic alloys has been presented in a recent article by Martin *et al.*,[11] which emphasizes the importance of the study at hand. MD simulations of the devitrification process[13] especially in the case of TFMG[13] have also been documented. Various recent reports on the theoretical basis and simulation studies of the devitrification process in metallic alloys,[11,12] including alloys of the Ni-Zr system,[14] highlight the significance of employing MD simulations to explain these phase transformations. Intermediate phases of the Ni-Zr system that are known to melt congruently,[15] such as $Ni_7Zr_2$, are ideal candidates for studying the amorphous to crystalline transformation. $Ni_7Zr_2$ possesses the highest melting point of all the known phases of this system and is, in addition, a mechanically stable phase.[16] This intermetallic compound is also reported to be the primary phase during solidification involving Ni-rich alloys[17,18] and Ni-based superalloys.[19] Its reportedly poor glass forming ability[20] provides an imperative need to investigate its glass forming ability and associated thermal stability in the thin film form. The formation and study of $Ni_7Zr_2$ TFMG have not been reported so far. Hence, this is the first study of this kind. An interesting aspect of a TFMG study pertains to the film thickness considered. The thermal stability with respect to crystallization aspects of TFMGs is intimately related to their thickness.[21] The thickness range studied for TFMGs corresponding to the Ni-Zr system is reported to be from 500 nm to 800 nm.[22,23] This thickness range becomes significant in a devitrification process study since a certain critical thickness of 100 nm, below which crystallization has not been observed to occur, has been defined through several studies of TFMGs of the Ni-Zr system.[24,25] The present work reports the formation of a Ni-Zr nanodimensional TFMG and its devitrification in a scaled down thickness regime. MD simulations applied to this devitrification process amply substantiate the experimental findings of this study.

## II. EXPERIMENTAL DETAILS

The Ni-Zr alloy films were deposited by DC magnetron cosputtering using independent Ni and Zr targets each with purity of ~99.95%, arranged in a confocal geometry below the substrate assembly. In this sputter-up configuration, the distance between the target and the substrate can be varied through vertical movement of the substrate manipulator, and during deposition, the substrates are rotated to ensure uniform deposition. Details of the sputter system used for this study have been reported elsewhere.[26] Sputter deposition parameters like power applied to the source targets, the source-substrate distance, and the sputter gas (Ar) pressure were optimized before each deposition. The base vacuum was $2.8 \times 10^{-7}$ Torr. Si wafers were used as substrates; they were chemically cleaned and mounted in the vacuum chamber prior to deposition through a load lock facility to minimize contamination in the system. Target-substrate distance was fixed at 75 mm. The Ar gas flow was tuned to 40 sccm through a mass flow controller maintaining 34 mTorr in the vacuum chamber during deposition. The targets were presputtered before use. Since atomic ratios of Ni/Zr can be controlled by varying the pulsed DC power settings for the two sputtering targets, the alloy films were formed on the substrates by cosputtering Ni and Zr targets while maintaining a DC power ratio of Ni:Zr as 0.44. Through this method, homogenous alloy films which were uniform both radially and throughout their thickness were formed. The alloy films were annealed at 800 °C in vacuum for 1 h. The films were characterized before and after annealing through various techniques for structural and magnetic properties. Each film thickness was estimated through model fitting of X-ray reflectivity data. The corresponding compositions of the binary alloys studied were determined using the methodology reported earlier[1] of a joint analysis of X-ray (XRR) and polarized neutron reflectivity (PNR) data for the same samples.

## III. RESULTS

The physical properties of Ni-Zr alloy thin films before and after annealing were characterized by various techniques as detailed below.

### A. Grazing incidence X-ray diffraction (GIXRD)

GIXRD measurements were made with a powder diffractometer using Cu $K_\alpha$ radiation (with wavelength $\lambda = 0.154$ nm). Diffraction patterns of as-deposited Ni-Zr alloy films recorded in a grazing geometry (incident angle 0.5°) showed an amorphous nature. However, the GIXRD scans of the 800 °C annealed films exhibited crystalline peaks which were further processed to identify the structure (space group, composition, etc.) using Le Bail fitting and Rietveld analysis methods. Figure 1 compares the structural difference between the as-deposited and the annealed films.

### 1. Le Bail fitting

The data were fitted using the Le Bail method, and the corresponding fit is shown in Fig. 1. From the fit, it could be deduced that the structure of the annealed Ni-Zr thin film was that of the $Ni_7Zr_2$ alloy phase.[27] The refined cell parameters obtained corresponding to the $C2/m$ space group ascertained were as follows: $a = 4.6731(4)$ Å, $b = 9.7455(4)$ Å, $c = 13.5061(4)$ Å, and $\beta = 96.803°$ (3). Initial parameters were taken from the Inorganic Crystal Structure Database (ICSD-2778). It was observed that the fitted lattice parameters were very close to that of a pure monoclinic structure:[27] $a = 4.698 \pm 0.009$ Å, $b = 8.235 \pm 0.013$ Å, $c = 12.193 \pm 0.016$ Å, and $\beta = 95.83 \pm 0.13°$. Thus, the annealed film crystallized into a slightly different structure, with the space group C2/m remaining intact. Observation of such a structure may be attributed to the large anisotropy introduced in the system due to thermal annealing leading to devitrification. The slight change in lattice parameters (atomic bonding) obtained after annealing is represented through the crystallographic structures shown in Fig. 2.

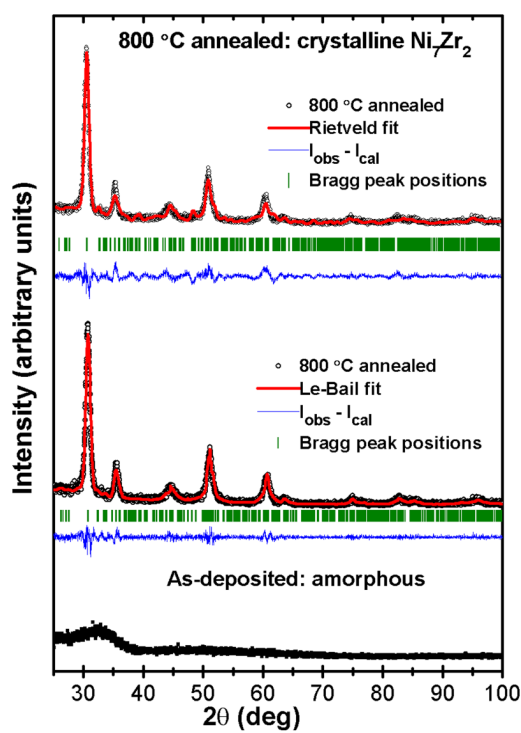

**FIG. 1.** GIXRD patterns of the as-deposited and the annealed Ni-Zr films. Fitting of annealed film data through the Rietveld and Le Bail methods revealed a crystalline structure corresponding to the $Ni_7Zr_2$ phase.

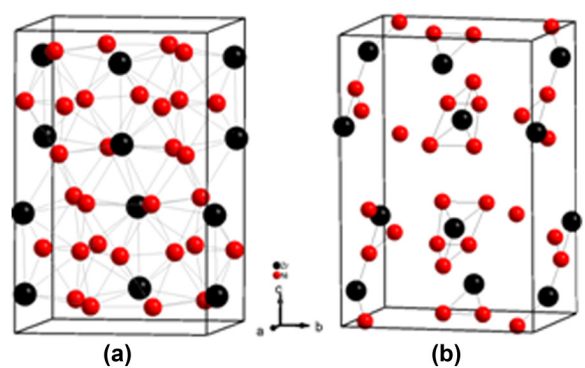

**FIG. 2.** Change in lattice parameters obtained for the annealed film with respect to the pure monoclinic structure is depicted through derived crystallographic structures: (a) pure structure and (b) annealed film crystallizes in *C2/m* structure.

### 2. Rietveld refinement

On the basis of confirmation of composition and other structural parameters from the Le Bail fit, the starting model of the $Ni_7Zr_2$ structure was used to refine the data to achieve a very reasonable fit between the observed data and the calculated pattern. Rietveld refinement profile parameters obtained were $\chi^2 = \sim 1$, $R_B = 5.65$, and Rf-factor = 3.43. Typical bond distances between Ni and Zr were calculated on the basis of structural refinement. For analyzing the GIXRD data, a peak shape profile of the Thompson–Cox–Hastings pseudo-Voigt type was chosen which consists of Lorentzian as well as Gaussian components. The average particle size (which represents the volume averaged diameter of crystallites in all directions) calculated from the Rietveld analysis of the GIXRD data was found to be around 4 nm. In addition, using the Debye–Scherrer formalism, the average crystallite size of the 800 °C annealed samples was found to be about 9.7 nm. The Rietveld analyzed structure of the annealed film is also included in Fig. 1.

## B. X-ray reflectivity (XRR) and polarized neutron reflectivity (PNR)

XRR data were recorded on the same instrument used for GIXRD measurements. XRR measurements were made on the Ni-Zr alloy thin films using parallel beam geometry and with a fixed slit size. This mode allowed the X-ray intensity to fall over a well determined area covering the entire sample. PNR data were recorded on the reflectometer used for various materials at the DHRUVA reactor, India.[28] The measurements were made with varying grazing angles of incidence $\theta$ on the sample surface, as a function of momentum transfer $q = (4\pi/\lambda)\sin\theta$ in Å$^{-1}$, so that reflected intensities were recorded in a perpendicular direction to the surface. Thus, the structural information could be obtained as a function of depth along the surface normal to the film. The XRR and PNR data recorded for as-deposited and annealed Ni-Zr alloy films were fitted using theoretical models, and the derived fitting parameters were used to quantitatively describe the physical structure of the films. Used in a specular configuration mode, XRR and PNR experiments probed the electronic and nuclear densities, respectively, which were averaged in the sample plane but varied along the depth of the sample. The associated depth-dependent scattering length density (SLD) in units of Å$^{-2}$, which is a function of number of atomic components per unit volume, for X-rays is given as electron SLD or $\rho_{e_i} = r_e \sum_i N_i Z_i$, where $r_e$ (=2.818 fm) is the classical electron radius and $N_i$ and $Z_i$ are the depth-dependent number density per unit volume and charge number, respectively, of the $i$th component of the alloy. Figure 3 shows the derived variation of electron SLD with depth for (a) as-deposited and (b) annealed films. The respective fitted XRR data are included as insets in Figs. 3(a) and 3(b).

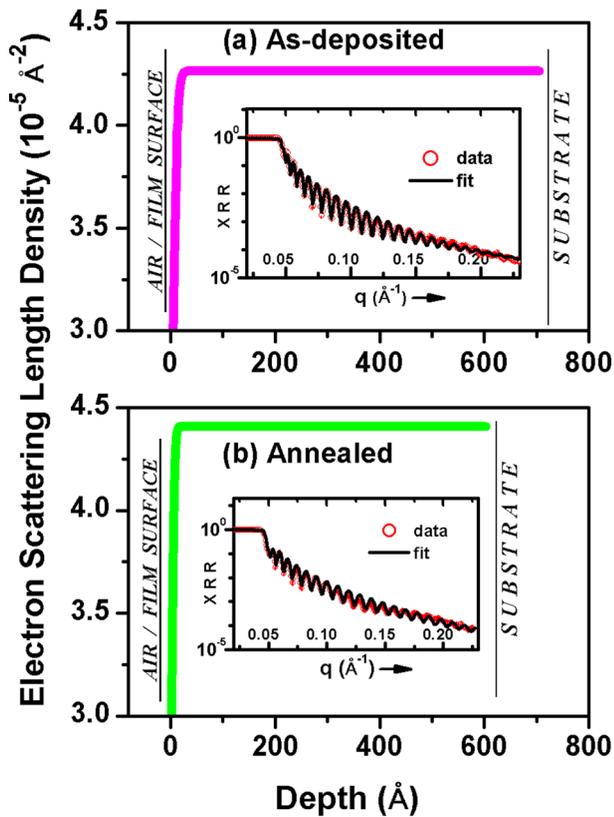

**FIG. 3.** Electron SLD depth profiles for (a) as-deposited (magenta solid line) and (b) annealed (green solid line) films plotted on the same scale, to highlight the differences in both SLD values and film thickness (depth). Insets are the corresponding fitted XRR data.

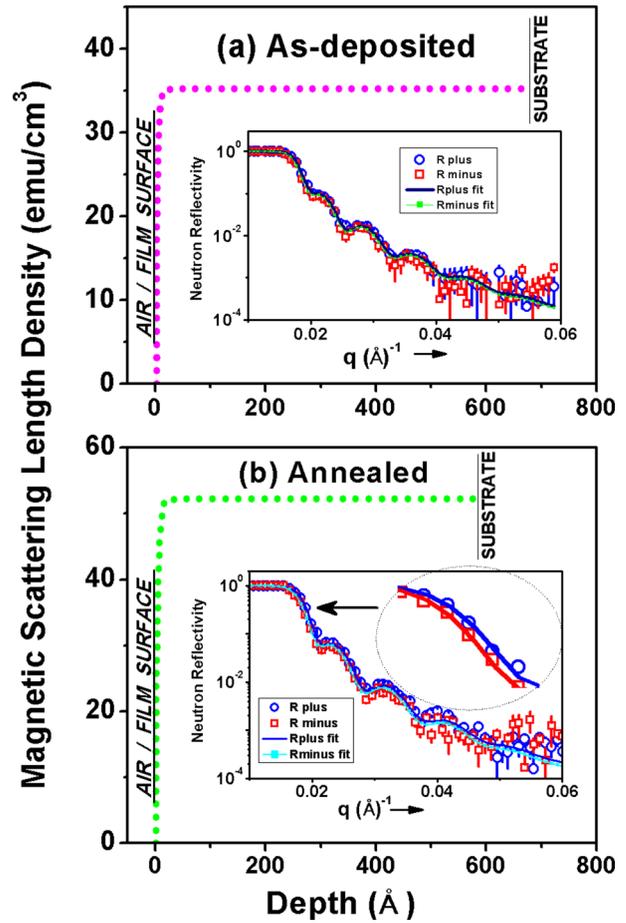

**FIG. 4.** Magnetic SLD depth profiles for (a) as-deposited (magenta dotted line) and (b) annealed (green dotted line) films. Ordinate values indicate the increased level of magnetic SLD after annealing. Insets are the corresponding fitted PNR data. Inset in (b) also highlights the separation between R+ and R− traces through a magnified part of the fitted data.

The neutron-nucleus interaction for any material is characterized by coherent neutron scattering length $b$, and the corresponding nuclear SLD extracted from PNR is given by the product $N_i b_i$, which can be used to determine chemical composition. The use of polarized neutrons offers the advantage of recording reflectivity R with neutron spin either parallel (R-plus) or antiparallel (R-minus) to the in-plane sample magnetization. PNR can detect small magnetic moments with superior depth resolution. The total SLD thus also includes a magnetic component and is given as $\rho_{n_i} = \sum N_i [b_i \pm C\mu_i]$, where $\mu_i$ is the magnetic moment of the $i$th element in Bohr magnetons which, therefore, has units of $\mu_B$/atom and $C = 2.69 \text{ fm}/\mu_B$ is a constant. Here, the neutron absorption coefficient is negligible and can be ignored. The $\pm$ sign depends on the orientation of sample magnetization with respect to the polarization direction of the neutron beam. Thus, magnetic moment density profile as a function of depth can also be determined from PNR. The variations of magnetic scattering length densities obtained by this method are shown in Fig. 4 for (a) as-deposited and (b) annealed films. The insets in these figures are the corresponding data and fits to R-plus and R-minus data.

A combination of XRR and PNR data analyses afforded a useful method to derive stoichiometry of binary alloy films.[29] In order to determine the atomic ratio of Ni:Zr in the films as a ratio $p:q$, the following set of linear equations were solved for both as-deposited and annealed cases:

$$\rho_n = pN_{Ni}b_{Ni} + qN_{Zr}b_{Zr},\tag{1}$$

$$\rho_e = pr_eN_{Ni}Z_{Ni} + qr_eN_{Zr}Z_{Zr},\tag{2}$$

where $\rho_n$ and $\rho_e$ are extracted from the respective fits to the PNR and XRR data of the binary alloy films. It may be noted that both these SLDs originate from a linear combination of the same number density $N$ (number of scatterers per unit volume) for each reacting element. The parameters obtained from independent fits to XRR and PNR data of as-deposited and annealed films, as well as from the joint analysis explained above, are compared in Table I.

TABLE I. Fit parameters derived from XRR and PNR data.

| Fit parameters | As-deposited | 800 °C annealed |
|---|---|---|
| Average film thickness (Å): joint analyses | 690 ± 10 | 590 ± 10 |
| Average film density (g/cc): joint analyses | 5.6 ± 0.05 | 5.8 ± 0.05 |
| Electron scattering length density (Å$^{-2}$): XRR fits | $4.2 \times 10^{-5}$ | $4.4 \times 10^{-5}$ |
| Magnetic scattering length density (Å$^{-2}$): PNR fits | $1.03 \times 10^{-7}$ | $1.52 \times 10^{-7}$ |
| Magnetization (emu/cc) from PNR fits | 34 | 50 |

The $p/q$ atomic ratio of Ni/Zr in both the as-deposited and 800 °C annealed samples was derived as 3.46. The atomic composition was computed to be Ni77.85Zr22.5 (at. %). This corresponds to the compound Ni$_7$Zr$_2$, which was also confirmed through grazing incidence XRD for the crystalline state. Atomic composition of the alloy films corresponding to the Ni$_7$Zr$_2$ single phase was thus preserved during preannealing and postannealing, indicating a polymorphous transition. Analyses of the fits revealed that the transformation of as-deposited Ni$_7$Zr$_2$ films to the annealed state resulted in a decrease in overall thickness (volumetric contraction) accompanied by an increase in average density. Percentage increase in density upon crystallization is ~3.6%, which is much higher than that reported for bulk amorphous alloys.[30,31] From PNR data analyses of the films measured under a field of 0.2 T, the magnetic moment corresponding to the as-deposited film was obtained as 0.07 $\mu_B$/atom and that for the annealed film was 0.1 $\mu_B$/atom. Thus, the magnetic moment increased ~1.4 times as a result of annealing. These values were converted to magnetization values (emu/cc) for easy comparison with those derived from superconducting quantum interference device (SQUID) measurements and are also listed in Table I along with other fit parameters.

In the case of amorphous structures, information regarding atomic arrangements is obtained from the radial distribution function (RDF), which provides the probability of neighboring atoms being located at varying distances from an average atom. MD simulations were effectively utilized in this work to compare the structure of amorphous and annealed films through estimating the extent of rearrangement of the nearest-neighbor shell for the structure with lowest energy.

## C. Computational methodology: MD simulations

MD simulations were performed to study the effect of annealing on amorphous Ni-Zr alloy film at the experimental composition of Ni$_7$Zr$_2$. In these MD simulations, the interactions between the atoms were described by the Finnis–Sinclair embedded atom method (EAM-FS) potential[32] usually used to describe transformation processes like devitrification.[33] The total energy $E_i$ of the "ith" atom in this formalism can be written as

$$E_i = F_\alpha \left( \sum_{j \neq 1} \rho_{\alpha\beta}(r_{ij}) \right) + \frac{1}{2} \sum_{j \neq 1} \varphi_{\alpha\beta}(r_{ij}), \qquad (3)$$

where the sites "$i$" and "$j$" at a distance "$r$" are occupied by $\alpha$ and $\beta$ type atoms, respectively. $F_\alpha(\rho)$ is the embedding energy which depends upon the electron density $\rho_{\alpha\beta}$, and $\varphi_{\alpha\beta}(r_{ij})$ represents the two body pairwise potential between atoms $i$ and $j$ separated by a distance $r_{ij}$. The EAM potential developed by Mendelev et al. was used to simulate the system under study.[33]

Large-scale atomic/molecular massively parallel simulator (LAMMPS) code[34] was used to carry out MD simulations. A simulation box containing $10 \times 10 \times 10$ unit cells of monoclinic Ni$_7$Zr$_2$ was taken as the initial structure to carry out the simulations. Three-dimensional periodic boundary conditions were used in the calculation to avoid surface effects. An isobaric-isothermal NPT ensemble (constant N: number of particles, P: pressure, and T: temperature) was employed using a Nose–Hoover barostat and a Nose–Hoover thermostat. The relaxation time for the barostat was 10 time units, while that for the thermostat was 0.1 time units. The system was equilibrated for 1 ns at a temperature of 4000 K, which was well above the melting temperature, to get a liquid state, and then the system was allowed to quench at a quenching rate of $10^{13}$ K/s to get the amorphous state. This amorphous structure was then annealed for 10 ns at 900 K in order to compare with the final desired crystalline structure. For the sake of completeness and to aid future modeling studies, the simulations were also performed with different annealing times: 5 ns and 50 ns; however, it was observed that the results were the same as that obtained for simulations with 10 ns. Finally, the radial distribution function (RDF) was calculated by binning pairwise distances into 500 bins from 0.0 to 7.0 Å which was less than the cut-off distance of the employed potential. Figure 5(a) shows plots of the obtained radial distribution functions for the liquid phase, the amorphous phase, the starting monoclinic phase, and the final annealed phase attained at 900 K.

The first radial distance of 3.5 Å considered for these simulations accommodates only the first peak of the RDF, whereas the second radial distance of 5.2 Å accommodates the first as well as the second peak of the RDF. The radial distribution functions of the liquid and the amorphous phases are quite similar as expected, with only one sharp peak for the first coordination shell and no correlations thereafter. In order to illustrate that the simulated amorphous state could indeed model the as-deposited TFMG, respective structure factor calculations were performed and are compared in Fig. 5(b). From Fig. 5(a), it can be seen that the annealed phase has almost fully transformed into the monoclinic phase from the amorphous structure. Detailed nearest neighbor configurations can offer better insight into the analyses of this amorphous to crystalline transformation. Table II lists the average coordination numbers for both Ni and Zr atoms in the initial amorphous state and the final crystalline state and at radial distances of 3.5 Å and 5.2 Å. The average coordination number of a given chemical species corresponds to the average number of nearest neighbors of any kind up to a certain radial distance. It is seen that the distribution of these numbers is consistent with the RDF analyses given above.

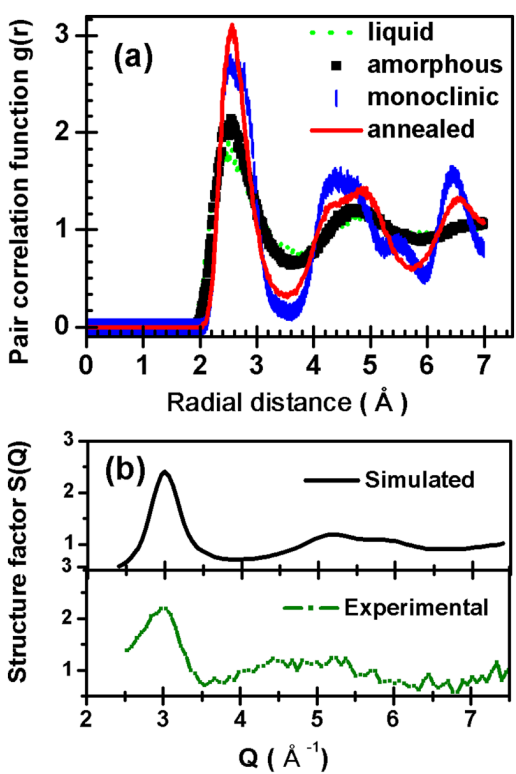

FIG. 5. (a) Radial distribution functions of the liquid phase, the amorphous phase, the starting monoclinic phase, and the transformed phase due to annealing at 900 K. (b) Comparison of computed structure factors for simulated and experimental amorphous states.

The local atomic configurations in the amorphous alloy are seen to be significantly different from those in the corresponding equilibrium crystalline phase. Structural rearrangements on an atomistic scale are defined in terms of short range order (SRO) in the system.[35]

In general, the average coordination number is seen to increase with annealing which indicates an increase in the extent of chemical SRO (CSRO), which defines the chemistry of intermolecular interactions. Investigating further beyond this overall trend, the coordination of unlike atoms greatly exceeds that of like atoms. This propensity toward pairing of unlike atoms over that of like atoms engenders both alloy formation and the transition to the crystalline state, as is seen in metallic glasses.[36] The increase in the Zr coordination environment of Ni atoms after annealing is seen to be equivalent to that of Ni around Zr atoms for both radial distances. The conclusions drawn from the RDF analyses lead to the inference that in this system, the strength of the heterogeneous Ni–Zr bond[37] greatly exceeds that of either Ni–Ni or Zr–Zr. This fact has been observed by other researchers too, especially for Ni-rich Ni-Zr alloys.[38] Here, in this case, it can be illustrated better by considering the potential of mean force between these atom pairs. To this end, the atom-resolved pair radial distribution functions were obtained through binning Ni-Zr distances, Ni-Ni distances, and Zr-Zr distances and are plotted in Fig. 6. These plots clearly indicate that in accordance with reversible work theorem, the Helmholtz free energy of the heterogeneous bond Ni–Zr is more negative and hence is stronger than the homogeneous bonds, viz., Ni–Ni and Zr–Zr. A strong tendency toward compound formation is thus indicated, aided by the highly negative value of enthalpy associated with alloy formation in the Ni-Zr system. The favorable tendency of compound formation and associated low glass forming ability of $Ni_2Zr_2$ have also been established by Li et al.[20] through the application of MD simulations to obtain certain thermodynamic properties of this system. In the initial state, the respective environments around Ni and Zr are similar since coordinate specificity is usually lost for amorphous structures as expected, which means that there is no distinction between coordination of Zr and Ni, unlike in the case of the crystalline state. Thus, a disorder to order transformation is stimulated in the Ni-Zr system as a result of annealing. On comparison of the percentage increase in nearest neighbors before and after annealing for the first radial distance, it is evident that the average coordination environment around Zr increases to a much larger extent than that about Ni. This effect may be attributed to the wide atomic size difference between Ni and Zr, which necessitates the long range rearrangement of Ni around Zr in the course of crystallization.

In order to further elucidate this aspect, Fig. 7 compares the simulated unit cell of a pure monoclinic structure with that of the annealed film. Red symbols correspond to the Ni atoms, while the Zr atoms are represented through blue symbols. The white portions of the symbols depicted in the simulation of the annealed film are indicative of the fraction of atoms below 100% occupancy. It was found that while the atom occupancy of Ni and Zr atoms in the pure monoclinic structure were both 100%, in the case of the annealed film, Ni atom occupancy was calculated to be 97.6% and that of Zr was 64%. This establishes the fact that the annealed sample is indeed a very close approximation to the monoclinic phase, with the lighter Ni atoms being rearranged to a larger extent than the heavier Zr atoms.

TABLE II. The average coordination numbers for Ni and Zr atoms in the initial amorphous and final crystalline phases at different radial distances and at 900 K. The bold numbers in parentheses are the respective average coordination numbers for a perfectly crystalline phase (monoclinic structure).

| Radial distance (Å) | Ni atoms around a Ni atom | | Zr atoms around a Ni atom | | Ni atoms around a Zr atom | | Zr atoms around a Zr atom | |
|---|---|---|---|---|---|---|---|---|
| | Initial amorphous | Final crystalline | Initial amorphous | Final crystalline | Initial amorphous | Final crystalline | Initial amorphous | Final crystalline |
| 3.5 | 8.26 | 8.63 (**8.29**) | 2.93 | 3.75 (**4.00**) | 10.26 | 13.13 (**14.00**) | 1.24 | 0.66 (**2.00**) |
| 5.2 | 28.75 | 34.26 (**35.14**) | 8.08 | 9.65 (**10.00**) | 28.29 | 33.76 (**35.00**) | 8.06 | 10.66 (**11.00**) |

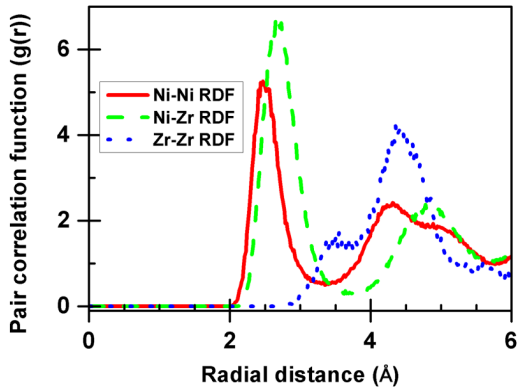

FIG. 6. Atom-resolved pair radial distribution function for the pairs Ni-Ni, Ni-Zr, and Zr-Zr.

## D. Atomic force microscopy (AFM)

The NT-MDT make P47H AFM system was used for morphological measurements on the as-deposited and annealed films. A cantilever probe of Si with length 125 $\mu$m, width 35 $\mu$m, and a super sharp diamondlike carbon (DLC) tip having a tip curvature

1–3 nm, resonance frequency 198 kHz, and force constant 8.8 N/m was used. The DLC AFM probe that was very hard and antiabrasive was chosen to ensure consistency during measurements. The surfaces of the as-deposited and annealed films were scanned over a $2 \times 2 \mu m^2$ area. A height-height correlation function given in Eq. (4) was computed from the AFM scanning measurements for each of the films. The exponential function usually applied to self-affine fractal surfaces is

$$H(r) = 2\sigma^2 \left[ 1 - e^{-(r/\xi)^{2\alpha}} \right], \qquad (4)$$

where "$\sigma$" is long order surface (RMS) roughness, "$\xi$" is the correlation length which is a measure of "geometric" grain size, and the dimensionless exponent "$\alpha$" which is referred to as the Hurst parameter is the short range roughness; in other words, it describes the grain surface structure and its value lies between 0 and 1. Suitable modeling of the function (4) enabled the derivation of these defining parameters. Figure 8 presents all the information extracted from this study, including data and fits of the height-height correlation function, AFM images, and height histogram curves obtained for both as-deposited and annealed films.

AFM image of the annealed alloy film exhibits the evolution of a well-defined, dense, and organized grain structure, whereas that of the as-deposited film depicts smoother surface morphology with very few irregular and disorganized islands possessing similar heights. The evolution of dense and organized nanograins on the surface of the annealed film is the characteristic attribute which qualifies the transition of amorphous to crystalline structure of the Ni-Zr alloy thin film. Correlation length and RMS roughness obtained for the as-deposited film were 39 nm and 5.3 Å, respectively, while the same for the annealed film were computed as 26.7 nm and 9.8 Å, respectively. Correlation length refers to the length scale value below which heights obtained are correlated.

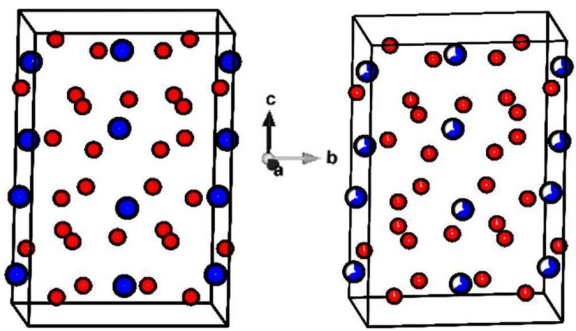

FIG. 7. The left panel contains the simulated unit cell of a pure monoclinic Ni$_7$Zr$_2$ structure to be used as a reference. The right panel shows the occupancy of Ni (red symbols) and Zr (blue symbols) atoms in this structure for the simulation of the annealed film. The white portion of the symbols is indicative of the fraction of atoms below 100% occupancy.

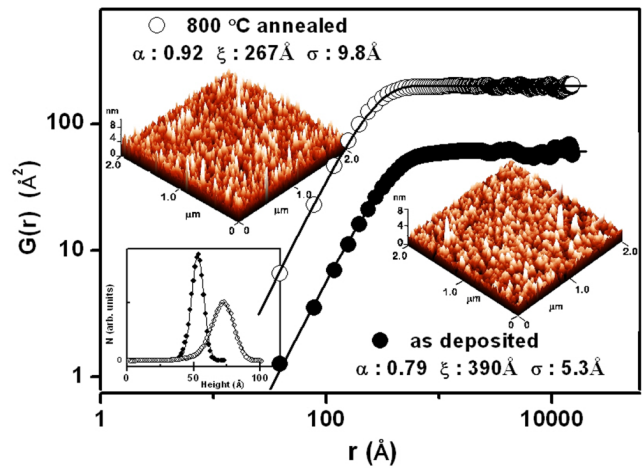

FIG. 8. Height-height correlation functions and height histograms derived from AFM data of as-deposited and annealed films.

**TABLE III.** Parameters obtained from AFM measurements.

| Parameter | As-deposited | Annealed |
|---|---|---|
| Surface RMS roughness $\sigma$ (Å) | 5.3 | 9.8 |
| Correlation length $\xi$ (nm) | 39 | 26.7 |
| Hurst $\alpha$ | 0.79 | 0.92 |
| Fractal dimension $D$ | 2.21 | 2.08 |

The annealed film thus depicts lower correlation length and higher RMS roughness as compared to the as-deposited film. Lower RMS values of roughness obtained for amorphous Ni-Zr alloy films have been reported earlier.[23] Surface roughness is affected by the growth mechanisms in films. Higher roughness obtained after annealing can be attributed to the surface nucleation of grains followed by columnar growth within the interior of the film. The value of the Hurst parameter is also seen to rise with increased temperature. The fractal dimension of the films can be described by the fractal geometry $D = 3 - \alpha$. This represents the degree to which the object under measurement occupies the space it is embedded in. Height histogram curves portray the quantitative characterization of the grain distributions. These histograms obtained from the AFM data of the two samples were each fitted with a Gaussian profile. Comparison of the same revealed a broader curve with a peak value at higher height for annealed films than the as-deposited film, wherein the surface showed a flatter morphology leading to a narrower height histogram peaking at lower height value. The parameters computed from the AFM measurements on the two films are listed in Table III.

### E. Magnetic measurements

Measurements to determine the magnetic nature of the films were carried out using a superconducting quantum interference device (SQUID) magnetometer (Quantum Design, model MPMS), with field applied parallel to the sample surface. Samples were sandwiched between pieces of a straw, and no separate sample holder was used. Temperature dependent magnetization [$M(T)$] measurements and field dependent magnetization [$M(H)$] recorded at 5 K and 300 K were made for both as-deposited and annealed films. SQUID data for the as-deposited and annealed films are shown in Fig. 9. The field dependent data have been fitted using a modified Langevin function:

$$M(H) = M_s[\coth(\mu H/kT) - (kT/\mu H)] + \chi_f H, \quad (5)$$

where $M_s$ is the saturated magnetization, $\mu$ is the average magnetic moment of the magnetic entity, and $\chi_f$ is the high field susceptibility of the sample showing a linear dependence on $H$.

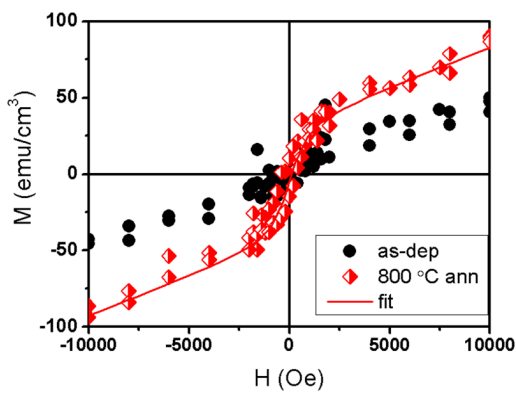

**FIG. 9.** SQUID data of the as-deposited and the annealed films. The fit to data from the annealed film is shown as a solid line.

The fit to Eq. (5) established two component magnetic responses of the sample, with a ferromagnetic component saturating at higher fields and a presumable superparamagnetic component showing a linear dependence with the field. Furthermore, the $M(H)$ curve of the annealed film showed a slight hysteresis at low fields indicative of a ferromagnetic nature associated with the long range order of atomic spins. The as-deposited data did not show any such feature reaffirming the disorder to order transition in the alloy films under discussion. The saturation magnetization values extracted from the fits were 30 emu/cc for the as-deposited and 48 emu/cc for the annealed films, which matches well with corresponding magnetic moment values derived from PNR data fits (Table I).

## IV. DISCUSSION

The atomic compositions of the cosputtered alloy films before and after annealing were deduced from a combined analysis of XRR and PNR data to be the same: Ni78Zr22. The as-deposited amorphous films had thus undergone a polymorphous transformation to a nanocrystalline state confirmed by GIXRD analysis to correspond to the Ni$_7$Zr$_2$ phase. The films were both homogeneous and uniform as observed from reflectivity measurements, and surface morphology could be related to the structural changes that the films had undergone, through AFM data analyses. AFM images revealed dense and uniform fine grains after annealing signifying that the films were highly crystallized. The grain structure evolution had led to a rougher surface morphology, as also seen through reflectivity analyses. The reflectivity data further indicated that an increase in density and shrinkage of thickness had accompanied the transformation. The ~3.6% rise in density as mentioned before in Sec. III B is much higher than that reported for bulk amorphous alloys. With regard to change in film thickness, earlier reports on devitrification of amorphous Ni-Zr alloy layers obtained within diffusion couples or multilayer interfaces define a critical thickness of 1000 Å below which crystallization did not occur. However, the thicknesses used in those studies ranged from 5000 Å to 8000 Å. In the present study, we have observed a thickness drop from 690 Å to 590 Å in the course of devitrification. The volume contraction is an expected consequence of densification of the metallic glass films after annealing to form the crystalline state. Annealing of glassy materials leads to relaxation to a more stable structure with lower free energy—an ordered arrangement such as a crystalline lattice. The atomic rearrangement during annealing leads to better packing, increased coordination number of atoms, and thereby densification through the decrease of free volume in the system.[39] This fact governs the structural relaxation process whereby the system is relieved of atomic level stresses. The above aspects become all the more significant during the kinetics of phase transformations in a nanodimensional thin solid film. As the film thickness is lowered, the atomic stresses increase, which restrain the progress of devitrification leading to a decreased rate of crystallization as compared to bulk glassy alloys. This feature of sluggish crystallization can be overcome only by annealing at elevated temperatures, as seen in metallic glasses. Annealing of TFMGs at high temperature can also provide insight into their thermal stability. The thermal stability of the amorphous phase in Ni-rich alloy thin films has also been attributed to the reduced enthalpy associated with the amorphous to crystalline transformation.[41] Alternate explanations for the link between thermal stability and film thickness of TFMGs of other materials have also been put forth.[42]

Differences between amorphous and crystalline phases are primarily described by the manner in which local structural units are connected or the short range structural difference. Although amorphous/glassy materials are usually classified as disordered, on an atomic scale, they are seen to constitute a local structural order.[43] It follows that the differences observed between these local units of crystalline and amorphous phases would thereby distinguish them from each other.[44,45] The amorphous (disorder) to crystalline (order) transition consists of a reduction in residual entropy of the system,[46] with an increase in the extent of CSRO. MD simulations applied to the transformation of the as-deposited amorphous Ni-Zr alloy to the annealed crystalline $Ni_7Zr_2$ alloy also showed an overall increase in the average coordination number as a result of annealing the system. In particular, the development of CSRO was dominated by the increase in the number of unlike pairs relative to others which implies a tendency to pair unlike atoms. The high densification observed also is effectively achieved by the increased CSRO. In case of Ni-Ni configuration, the coordination around Ni atoms increased after annealing. Ni atoms being arranged closer to each other can be the reason for the larger magnetic moment of the alloy obtained in the annealed state, as also observed by others after devitrification in such systems.[47]

## V. CONCLUSIONS

Amorphous (glassy) to (nano) crystalline transformation of a thin cosputtered Ni-Zr alloy film, with thickness lower than a reported critical limit of devitrification, has been achieved and examined thoroughly. Whether the limit achieved through this work is critical to the particular $Ni_7Zr_2$ composition or not can be the subject of future study. Amorphous Ni-Zr alloy thin films were subjected to devitrification at 800 °C to form a nanocrystalline alloy film. The alloy compositions before and after annealing were deduced to have an atomic ratio of Ni/Zr as 3.46 through reflectivity analyses. The films had thus undergone a polymorphous transformation. This atomic ratio corresponds to the compound $Ni_7Zr_2$, which was confirmed from the Rietveld analysis of GIXRD data for the annealed crystalline monophasic state. Crystallite size of the order of 9.7 nm was estimated for the same. The thin films formed were homogenous and uniform, and their respective structures could be correlated with the surface morphology through AFM analyses. Devitrification was achieved in a nanodimensional regime, in amorphous thin films of 69 nm thickness. Another attribute of the devitrified nanofilm was the high percentage change in density, associated with structural relaxation in metallic glasses. Densification of the alloy films after annealing was linked to the increased short range order in the system through MD simulations. The changes obtained in magnetic properties after annealing the thin alloy films could also be explained adequately through MD simulations applied to this process. The increase in magnetic moment after devitrification was suitably correlated with the simulated change in the Ni coordination number around Ni atoms. MD simulations further established that the favorable tendency toward pairing of unlike atoms was effective to cause structural relaxation during devitrification. Structural relaxation implies that the coordination number increases by densification. MD simulations could thus describe the devitrification of amorphous Ni-Zr alloy films to a stable nanocrystalline state to be a disorder to order transformation.

## ACKNOWLEDGMENTS

The authors would like to thank S. Jana and P. Moundekar for technical assistance; P. U. Sastry and V. B. Jayakrishnan for the use of XRR/GIXRD facility; and finally, all the reviewers for their useful criticism to enrich the manuscript favorably.